\documentclass{article}

\begin{document}
\title{High Order Relativistic Corrections To Keplerian Motion}
\author{L. Fern\'andez-Jambrina\\
Departamento de Ense\~nanzas B\'asicas de la Ingenier\'{\i}a Naval\\
E.T.S.I. Navales\\ Arco de la Victoria s/n\\ E-28040-Madrid, Spain \\ and 
\\
C. Hoenselaers\\ Department of Mathematical
Sciences\\ Loughborough University of Technology\\ Loughborough LE11
3TU, United Kingdom}
\date{}

\maketitle
\begin{center}PACS numbers: 04.25 Nx, 95.10.Ce
\end{center}

\begin{abstract}
    The first terms of the general solution for an asymptotically 
flat stationary axisymmetric vacuum spacetime endowed with an equatorial 
symmetry plane are calculated from the corresponding Ernst potential up to 
seventh order in the radial pseudospherical coordinate. The metric is  used
to determine the influence of high order multipoles in the perihelion 
precession of an equatorial orbit and in the node line precession of a 
non-equatorial orbit with respect to a geodesic circle. Both results are 
written in terms of invariant quantities such as the Geroch-Hansen 
multipoles and the energy and angular momentum of the orbit. 
\end{abstract}

\newpage

\section{Introduction} The construction of new exact solutions of the 
Einstein vacuum field equations describing an asymptotically flat
stationary  axially symmetric spacetime has increased in the last decade
due to the  several methods of generation of solutions from a given one,
such as the  B\"acklund transformations, the inverse scattering method and
the HKX  tranformations (cf. for instance \cite{Dietz} and references
quoted therein.)  Nevertheless we are very far from being able to implement
the physical  behaviour of an exact solution at will and in order to obtain
results which  could be tested experimentally we are led to use approximate
expressions for  the metrics with the desired physical requirements.

In this paper we study the influence of the first Geroch-Hansen multipole
moments  \cite{Geroch},\cite{Hansen} of order higher than three in some
astrophysical  situations with stationary and axial symmetry. As it is well
known, these  moments can be calculated from the coefficients of the power
expansion of  the Ernst potential on the symmetry axis \cite{Fodor}, being
linear the relation between both families of constants  until the third
moment, that is, the octupole. The subsequent expressions for  the
multipole moments of order higher than three become considerably more 
complicated as the order increases and there is not even a closed formula 
for calculating all of them. The purpose of this paper is therefore to show 
to what extent these nonlinearities affect the motion of test particles 
tracing their orbits around a non-spherical, in principle rotating, compact 
mass distribution. Of course, these terms are irrelevant for our solar
system  calculations, but they are meaningful for highly relativistic
astrophysical  objects, such as pulsars, as stated in \cite{Schafer}.

With this aim in mind we calculate in section \ref{metric} the first seven
terms of a power expansion of the Ernst potential with arbitrary values for
the multipole moments and construct the corresponding approximate metric.
This result is used in section \ref{precession} to obtain an expression for
the perihelion precession of an equatorial trajectory and in section
\ref{node} to produce the corrections to the Newtonian precession of the
nodes of a slightly non-equatorial orbit with reference to a close
neighbouring geodesic circle. There is some previous work on this subject
in \cite{quadr} and \cite{Quevedo}, but these references deal only with
terms up to the quadrupole moment. The results will be discussed in section
\ref{discussion}. 

\section{Calculation of the metric\label{metric}} 

The metric of a stationary axially symmetric vacuum spacetime can be
written in a canonical form in terms of the Weyl coordinates,

\begin{equation}
ds^2=-f(dt-Ad\phi)^2+\frac{1}{f}\{e^{2\gamma}(d\rho^2+dz^2)+\rho^2\,d\phi^2\},
\end{equation} where $t$ and $\phi$ are the coordinates associated with the
commuting Killing vectors $\partial_t$ and $\partial_\phi$, and the
functions $f$, $A$ and $\gamma$ depend only on the coordinates $\rho$ and
$z$. 

The whole set of Einstein equations can be shown \cite{Ernst} to be
equivalent to the following system of partial differential equations,

\begin{equation} \varepsilon=f+i\chi \end{equation}

\begin{equation}
\varepsilon_{\rho\rho}+\frac{1}{\rho}\varepsilon_\rho+\varepsilon_{zz}=
\frac{2}{\varepsilon+\bar\varepsilon}({\varepsilon_\rho}^2+{\varepsilon_z}^2)
\label{ernst}\end{equation}

\begin{equation}
A_\rho=\frac{4\,\rho}{(\varepsilon+\bar\varepsilon)^2}\,\chi_z\label{a1}
\end{equation}

\begin{equation}
A_z=-\frac{4\,\rho}{(\varepsilon+\bar\varepsilon)^2}\,\chi_\rho\label{a2}
\end{equation}

\begin{equation}
\gamma_\rho=\frac{\rho}{(\varepsilon+\bar\varepsilon)^2}\,(\varepsilon_\rho\bar
\varepsilon_\rho-\varepsilon_z\bar\varepsilon_z)\label{g1} \end{equation}

\begin{equation}
\gamma_z=\frac{\rho}{(\varepsilon+\bar\varepsilon)^2}\,(\varepsilon_\rho\bar
\varepsilon_z+\varepsilon_z\bar\varepsilon_\rho).\label{g2} \end{equation}

It can be shown that the integrability of the last four equations is
guaranteed if equation (\ref{ernst}), the Ernst equation, is satisfied.
Therefore, in order to obtain a solution of the Einstein equations with two
Killing vectors, it suffices to solve the Ernst equation and then calculate
the metric functions by quadratures.

It is also usual to write the Ernst equation in terms of another potential,
$\xi$, related to the previous one by the following relation,

\begin{equation} \xi=\frac{1-\varepsilon}{1+\varepsilon}, \end{equation}
which satisfies another partial differential equation, 

\begin{equation}
(\xi\bar\xi-1)\,(\xi_{\rho\rho}+\frac{1}{\rho}\xi_\rho+\xi_{zz})=
2\,\bar\xi\,({\xi_\rho}^2+{\xi_z}^2), \label{ernst2} 
\end{equation}which is also satisfied by $\xi^{-1}$. This latter 
form is the one introduced originally in \cite{Ernst} and allows a 
simple integration of the Kerr metric.

This form is particularly useful for calculating the multipole moments and
it will be the one we shall employ.

In order to calculate the approximate solution, we shall write the Ernst
potential $\xi$ as a function of two coordinates $r$ and $\theta$ related
to the Weyl ones by the following transformation,

\begin{equation} \rho=r\,\sin\theta\hspace{2cm} z=r\,\cos\theta.
\end{equation}

We can implement the requirement of asymptotic flatness by writing $\xi$ as
a formal inverse power expansion in the pseudospherical radial coordinate
$r$,

\begin{eqnarray} \xi=\sum_{n=1}^{\infty}\xi_n r^{-n}, \end{eqnarray} where
the functions $\xi_n$ depend only on the coordinate $\theta$.

Since we are interested in having a solution which is symmetric with 
respect to the equatorial plane $\theta=\pi/2$, we shall require that the 
functions $\xi_n$ of odd order be real whereas those of even order will  be
taken to be imaginary. 

With this information at hand we can now proceed to calculate the metric
functions. The function $f$ is just the real part of $\varepsilon$ and from
our knowledge of $\xi$ we can calculate the first eight  terms of its
expansion in the radial coordinate,

\begin{eqnarray} f=1+\sum_{n=1}^{\infty}f_n r^{-n} \end{eqnarray}

\begin{eqnarray} f_1=-2\,{ m_0} \end{eqnarray}

\begin{eqnarray} f_2=2\, { m_0}^{2} \end{eqnarray}

\begin{eqnarray} f_3={ m_2}-\frac {4\,{ m_0}^{3}}{3}
-3\,m_2\,\cos^2\theta\end{eqnarray}

\begin{eqnarray} 
{ f_4}=-2\,{ m_0}\,{ m_2}+{\frac {2\,{ m_0}^{4}}{3}}+
\left (6\,{ m_0}\,{ m_2}+2\,{ m_1}^{2}\right
)\cos^2\theta 
\end{eqnarray}

 \begin{eqnarray} 
 { f_5}&=&-{\frac {3\,{ m_4}}{4}}+2\, { m_0}^{2}{ m_2}+{\frac {8\,{ m_0}\,{
 m_1}^{2}}{35}}-{\frac {4\,{
 m_0}^{5}} {15}}+\nonumber\\ &+& \left ({\frac {15\,{ m_4}}{2}}-6\,{ m_0}^{2}{
 m_2}-{\frac {44\,{ m_0}\,{ m_1}^{2}}{7}}\right )\cos^2\theta-{\frac {35\, {
 m_4}\,\cos^4\theta}{4}}
 \end{eqnarray}


\begin{eqnarray} 
{ f_6}&=& {\frac {3\,{ m_0}\,{ 
m_4}}{2}}+{\frac
{{ m_2}^{2 }}{2}}-{\frac {4\,{ m_2}\,{ m_0}^{3}}{3}}-{\frac {16\,{ m_0}^{2}{ m_1}^{2}}{35}}
+{\frac {4\,{ m_0}^{6}}{45}}+ \nonumber\\&+&\left (-15\,{ m_0}\,{ m_4}-6\,{ 
m_1}\,{ m_3}-3\,{ m_2}^{2}+4\,{ m_2}\,{
m_0}^{3}+{ \frac {356\,{ m_0}^{2}{ m_1}^{2}}{35}}\right
)\cos^2 \theta+\nonumber\\&+&\left ({\frac {35\,{ m_0}\,{
m_4 }}{2}}+10\,{ m_1}\,{ m_3}+{\frac {9\,{ m_2}^{2}}{2}}\right )\cos^4\theta
\end{eqnarray}

\begin{eqnarray} 
{ f_7}&=&{\frac{5\,{  m_6}}{8}}-{\frac {3\,{ m_0} ^{2}{ m_4}}{2}}-{\frac {4\,{
m_0}\,{ m_1}\,{ m_3}}{11}}-{ m_0}\,{ m_2}^{2}-{\frac {16\,{ m_1 }^{2}\,{
m_2}}{231}}+\nonumber\\
&+&{\frac {2\,{ m_0}^{4}{ m_2}}{3}}+{\frac {32\,{ m_0}^{3}{ m_1}^{2}}{63}}
-{\frac {8\,{ m_0}^{7}}{315}}
+\nonumber\\&+&
\left (-{\frac { 105\,{ m_6}}{8}}+15\,{ m_0 }^{2}{ m_4}+{\frac {216\,{ m_0}\,{ m_1}\,{
m_3}}{11}}+6\,{ m_0}\,{ 
m_2}^{2}+\right.\nonumber\\ &+&\left.{\frac {38\,{ m_1}^{2}\,{ m_2}}{11}}
-2\,{ m_0 }^{4}{
m_2} -
{\frac {1208\,{ m_0}^{3}{ m_1}^{2}}{105}}\right)\cos^2\theta+\nonumber\\&+&
\left ({\frac {315\,{  m_6}}{8}}-{\frac {35\,{ m_0}^{2}{
m_4}}{2}}-{\frac
{340\,{ m_0}\,{ m_1}\,{ m_3}}{11}}-{\frac {114\,{
m_1}^{2}\,{ m_2}}{11}}-\right.\nonumber\\ &-&\left.9\,{ m_0}\,{ m_2}^{2}\right ) \cos^4\theta-{\frac {231\,{
m_6}}{8}}\,\cos^6\theta,
\end{eqnarray}
where the 
constants $m_n$ which arise from the integration of the Ernst equation are
real if $n$ is even and otherwise imaginary. 

The first six terms of the metric function $A$ can be obtained by direct 
integration of the equations (\ref{a1}) and (\ref{a2}). In spite of the 
factor $i$ before the expression for $A$, this function is obviously real
since  the constants $m_n$ are imaginary for odd $n$,

\begin{eqnarray} A=-i\sin^2\theta\sum_{n=1}^{\infty}A_nr^{-n} \end{eqnarray}

\begin{equation} A_1=-{2\,{m_1}} \end{equation}

\begin{equation} A_2= -{2\,{ m_0}\,{ m_1}} \end{equation}

\begin{equation} A_3= { m_3}-{\frac {8\,{
m_0}^{2}{m_1}}{5}} -5\,{ m_3}\,\cos^2\theta\end{equation}

\begin{eqnarray}
A_4&=&\frac{3\,m_0\,m_3}{2}+\frac{m_1\,m_2}{2}-\frac{16\,{m_0}^3\,m_1}{15}+
\nonumber\\&+& \left(-\frac{15\,m_0\,m_3}{2}+\frac{3\,m_1\,m_2}{2}\right)
\cos^2\theta\end{eqnarray}

\begin{eqnarray} 
A_5&=&-\frac {3\, 
m_5}{4}+
\frac {4\, {m_0}^{2}\,m_3}{3}+ \frac {8\, m_0\, m_1\, m_2}{7}
-\frac {64\, {m_0}^4\, m_1}{ 105}+ \nonumber\\ &+&
\left(\frac{21\,m_5}{2}-\frac {20\, {m_0}^2 m_3}{3}\right)\cos^2\theta-
\frac {63\, m_5}{4}\,\cos^4\theta
\end{eqnarray}

\begin{eqnarray} 
A_6&=&-{ \frac
{5\,{ m_0}\,{ m_5}}{4}}-{\frac {{ m_1}\, { m_4}}{4}}
-{\frac {{ m_2}\, { m_3}}{2}} +{\frac {8\,{ m_0}^{3}\,{ m_3}}{9 }}+\nonumber\\&+&
{\frac {48\,{ m_0}^{2}\,{m_1}{ m_2}}{35}} +{\frac {8\,{ m_0}{ m_1}^{3}}{105}}-{\frac {32\,{
m_0}^{5}\,{ m_1}}{105}}+
\left(\frac {35\, m_0\, m_5\,}{2
}- \frac {5\,m_1\,m_4}{2}+\right.\nonumber\\ &+&\left.m_2\,m_3-\frac {40
\, {m_0}^{3}\, m_3}{9}-\frac {8\, {m_0}^{2}\, m_1\, m_2 }{ 5}+\frac
{16\,m_0{m_1}^3}{21}\right)\cos^2\theta +\nonumber\\ &+&\left(-\frac {105\, m_0\, m_5\,}{4}+\frac {35 \, m_1\, m_4\,}{4}-\frac {5\,m_2
\,m_3}{2}\right)\cos^4\theta.
\end{eqnarray}

The only function which remains to be calculated is $\gamma$ and can be 
obtained as a quadrature from equations (\ref{g1}) and (\ref{g2}),

\begin{eqnarray} \gamma=\sum_{n=1}^{\infty}\gamma_{n}r^{-2n} \end{eqnarray}

\begin{eqnarray} \gamma_1=-\frac{{m_0}^2\,\sin^2\theta}{2} \end{eqnarray}

\begin{eqnarray} \gamma_2&=&{\frac {3\,{ m_0}\,{ m_2} }{4}}-{\frac {{ m_1}^{2}}{4}}
+\left(-{\frac {9\,{ m_0 }\,{ m_2}}{2}+ {\frac {5\,{ 
m_1}^{2}}{2}}}\right)\cos^2\theta +\nonumber\\&+&\left({\frac { 15\,{ m_0}\,{ m_2}}{4}}-{\frac
{9\,{ m_1}^{2}}{4}}\right)\cos^4\theta
\end{eqnarray}

\begin{eqnarray} 
\gamma_3&=&
-{ \frac {5\,{ m_0}\,{ m_4}}{8}}+{\frac {{ m_1}\,{ m_3}}{2}} 
-{\frac{3\,{ m_2}^{2}}{8}}+{\frac
{2\,{ m_0}^{2} { m_1}^{2}} {35}}
+ \nonumber\\ &+&\left({ \frac {75\,{ m_0}\,{
m_4}\,}{8}}-{\frac {21\,{ m_1}\,{ m_3}}{2}}+{\frac {4 5\,{
m_2}^{ 2}}{8}}-{\frac {2\,{ m_0}^{2}{
m_1}^{2}}{35}}\right)\cos^2\theta
+ \nonumber\\ &+&
\left(-{\frac
{175\,{ m_0}\,{ m_4}\, }{8}}+{\frac {55\,{ m_1}\,{
m_3}}{2}}-{\frac {117\,{ m_2}^{2}}{8}} \right)\cos^4\theta 
+ \nonumber\\ &+&
\left({\frac
{105\,{ m_0}\,{  m_4}}{8}}-{ \frac {35\,{ m_1}\,{
m_3}}{2}}+{\frac {7 5\,{ m_2}^{ 2}}{8}}\right)\cos^6\theta
\end{eqnarray}

\begin{eqnarray} 
\gamma_4&=&{\frac {35\,{ m_0}\,{
m_6}}{64}}-{\frac {15\,{ m_1}\,{ m_5}}{32}}+{\frac {45\,{ m_2}\,{ m_4}}{64}}-
\nonumber\\ &-&
{\frac {9\,{
m_3}^{2}}{32}}-{\frac {14\,{
m_0}^{2 }{ m_1}\,{ m_3}}{165}}-{\frac {2\,{ m_0}\,{ m_1}^{2}{ m_2}}{33}}+{\frac {16\,{ m_0}^{4}{ m_1}^{2}}{1575}
}+ \nonumber \\&+&
\left (-{\frac {245\,{m_0}\,{ m_6}}{16}}+{\frac
{135\,{ m_1}\,{m_5}}{8}}-{\frac {315\,{
m_2}\,{ m_4}}{16}}+{\frac {81\,{ m_3}^{2}}{ 8}}+
\right.\nonumber\\&+&\left. {\frac
{28\,{ m_0}^{2}{m_1}\,{m_3}}{55}}-{\frac {4\,{ m_0}\,{ m_1}^{2}{ 
m_2}}{231}}+{\frac{16\,{ m_0}^{4}{ m_1}^{2}}{1575}}\right )\cos^2\theta+\nonumber\\&+&
\left(\frac {2205\,{ m_0}\,{ m_6}}{32}-{\frac {1365\,{ m_1}\,{m_5}}{16}}+{\frac
{3075\,{ m_2}\,{ m_4}}{32}} -\right.\nonumber\\&-& \left.
{\frac {795\,{ m_3}^{2}}{16}}-{\frac {14\,{ m_0}^{2}{ m_1}\,{
m_3}}{33}}+ {\frac {6\,{ m_0}\,{ m_1}^{2}{
m_2}}{77}}\right ) \cos^4\theta+\nonumber\\&+&\left(
-{\frac {1617\,{
m_0}\,{m_6}}{16}}+{\frac{1071\,{ m_1}\,{
m_5}}{8}}-{\frac {2415\,{ m_2}\,{ m_4}}{16}}+{\frac{625\,{m_3}^{2}}{
8}}\right )\cos^6\theta+\nonumber\\&+&\left (
{\frac {3003\,{ m_0}\, { m_6}}{64}}-
{\frac {2079\,{ m_1} \,{ m_5}}{32}}+{\frac {4725\,{ m_2}\,{
m_4}}{64}}-{\frac {1225\,{
m_3}^{2}}{32}}\right)\cos^8\theta.
\end{eqnarray}

\section[Perihelion precession]{Perihelion precession of a closed orbit\label{precession}} 

It is well known from Bertrand's theorem (cf. for instance
\cite{Goldstein})  that stable bounded orbits of particles moving under the
influence of a  central force which is neither Newtonian nor harmonic are
not closed.  Therefore whenever the source of the gravitational field is
not exactly  monopolar, the bounded trajectories on the equatorial plane
will no longer  be the elliptic orbits described by Kepler's first law but
will take the form  of a precessing ellipse if the deviation from spherical
symmetry is small. 

The situation becomes a bit more complicated in general relativity.
Although  it has been proven \cite{Perlick} that there are just two
asymptotically flat  static spherically symmetric spacetimes in which the
stable orbits are  closed, they are rather different from the classical
physical situations.  Therefore, when relativistic effects are taken into
account, not even the  motion around a spherical distribution of mass is
closed. This effect has  been tested in our solar system and it amounts to
a slow precession of the  perihelion of the orbit of Mercury. Of course
other multipole moments of the  mass distribution will also contribute to
this effect and, in principle,  these moments could be calculated by
measuring the precession of a certain  number of test particles orbiting at
conveniently different distances from  the gravitational source. 

If the test particles are small enough for the tidal forces to be
unimportant within the characteristic length of the particle, we can  regard
them as point particles. If the effects due to their intrinsic  angular
momentum  can be taken as negligible it can be assumed that they trace out 
timelike geodesics in the spacetime surrounding  the gravitational source.
Hence, in order to study the influence of the  far field multipole moments
of the gravitational field, we shall have to  solve the geodesic equations
for the previously calculated metric. We shall  restrict ourselves to the
equatorial plane $\theta=\pi/2$.

Since the timelike and azimuthal coordinates are ignorable, we have two 
first integrals for the motion corresponding to the conserved quantities 
$E$ and $l$, respectively the total energy per unit of mass and the 
projection of the angular momentum on the $z$ axis, also per unit of mass, 
of the test particle. In terms of its 4-velocity  $u=(\dot t,\dot r,
\dot\theta,\dot\phi)$ these quantities have the  following form,

\begin{equation} E=-\partial_t\cdot u=f\,(\dot t-A\dot\phi) \end{equation}

\begin{equation} l=\partial_\phi\cdot u=f\,A\,(\dot
t-A\dot\phi)+\frac{1}{f}\,r^2\,\dot\phi, \end{equation} 
where the overhead dot stands for the derivative with respect to proper
time.

Therefore the equations for $t$ and $\phi$ can be written as follows,

\begin{equation} \dot\phi=f\,\frac{l-E\,A}{r^2} \label{phi} \end{equation}

\begin{equation} \dot t=\frac{E}{f}+f\,A\,\frac{l-E\,A}{r^2}. \label{t}
\end{equation}

Another integral arises from the fact that the trajectory is timelike and 
therefore $u\cdot u=-1$. For the geodesics under consideration this means, 

\begin{equation} -1=-f(\dot t-A\dot\phi)^2+\frac{1}{f}(e^{2\gamma}\,\dot
r^2+r^2\,\dot\phi^2).\label{timelike} \end{equation}

From the previous three equations $\dot r$ can be obtained as a function of 
the  non-ignorable coordinates and the conserved quantities. However, since
we  are interested in the shape of the orbit rather than in its time
evolution,  we divide (\ref{timelike}) by $\dot\phi$ to get the  derivative
of the radial coordinate with respect to the azimuthal angle, 

\begin{equation}
{r_\phi}^2=e^{-2\gamma}\left\{\frac{r^4\,(E^2-f)}{f^2\,(l-E\,A)^2}-r^2\right\}.
\end{equation}

It will be useful to write this equation in terms of another function 
$u=1/r$ as it is done in classical mechanics for solving the motion under 
central forces, 

\begin{equation}
{u_\phi}^2=e^{-2\gamma}\left\{\frac{E^2-f}{f^2\,(l-E\,A)^2}-u^2\right\}=F(u)=
\sum_{n=0}^{6}c_n\,u^n+O(u^7). \label{binet1} \end{equation}

This equation can be turned into a quasilinear one by taking a derivative 
with respect to $\phi$ and cancelling the $u_\phi$ factors, since for the
analysis of  perihelion precession circular orbits are of no interest,

\begin{equation} u_{\phi\phi}=\frac{1}{2}\,F'(u).\label{binet2}
\end{equation}

In order to solve these equations perturbatively we need expand them in 
powers of a small parameter. A good candidate is the inverse of the angular 
momentum per unit of mass, $l$, since according to Kepler's 1-2-3 law, 
which is assumed to be a good approximation at a great distance from the 
source, it behaves as $l\sim\sqrt{m\,r}$, where $m$ is the mass of the 
particle. It can be combined with the mass of the source $m_0$ to yield  an
acceptable dimensionless small parameter for analysing the far 
gravitational field. Hence we shall use $\epsilon=m_0/l$ and expand $u$  in
the following way, 

\begin{equation}
u=\epsilon^2\,\sum_{n=0}^{11}u_n\,\epsilon^n+O(\epsilon^{14}). \end{equation}

The reason for starting the expansion at this order is that the expression 
for the Kepler ellipse, which is expected to be the first term, is second 
order in $\epsilon$.

The energy per unit of mass of the particle is also to be expanded in 
$\epsilon$. Therefore we write,

\begin{equation} E=1+\epsilon^2\,\sum_{n=0}^{n=11}
E_n\,\epsilon^n+O(\epsilon^{14}). \end{equation}

In order to avoid the appearance of secular terms we use a coordinate
$\psi$  related to $\phi$ by,

\begin{equation}
\psi=\omega\phi\hspace{1.5cm}\omega=\sqrt{1+\sum\omega_i\epsilon^i}.
\end{equation}

The coefficients $c_n$ are all of the order $\epsilon^2$ except $c_2$ which 
is clearly of zeroth order in $\epsilon$ and therefore equation 
(\ref{binet1}) takes the form of a hierarchy of forced harmonic oscillators 
which can be solved iteratively up to the order of accuracy provided by our 
knowledge of the metric,

\begin{equation} u_{n\,\psi\psi}+u_{n}=f_n(\psi). \end{equation}

The first terms of the expansion of the solution to the equations 
(\ref{binet1}) and (\ref{binet2}) are,

\begin{equation}
u_0=\frac{1}{m_0}(1+\sqrt{1+2\,E_0}\cos\psi)\label{Kepler0} \end{equation}

\begin{equation} u_1=0 \end{equation}

\begin{equation} u_2=\frac{6+4\,E_0}{m_0} \end{equation}

\begin{equation} u_3=(8+4\,E_0)\,\frac{i\,m_1}{m_0^3}. \end{equation}

Of course the term of lowest order is the Kepler ellipse if $E_0$ is
negative. 

From the 
information we have about the metric we can calculate $\omega$ up to the 
eleventh power of $\epsilon$. These terms are just what we need to 
calculate the expression for the perihelion precession, so we shall 
focus on them. The expressions of the terms $u_{n}$ are not needed 
and therefore we shall no enclose them here. 

Instead of writing the results as a function
of  the integration constants $m_i$, it will be more useful to write them in
terms  of the Geroch-Hansen multipole moments, $P_i$, the physical
interpretation  of which is more appealing. Bear in mind that the odd
multipole moments are  imaginary  and have to be multiplied by $-i$ to
obtain the usual real expressions $J_n$.  To calculate these moments we
shall make use of the procedure described in  \cite{Fodor}. 

If we have and expansion of the Ernst potential $\xi$ on the symmetry axis 
in terms of the Weyl coordinate $z$, viz.

\begin{equation} \xi(\rho=0)=\sum_{n=0}^\infty C_n\,z^{-(n+1)} \end{equation} 
then the multipole moments can be calculated as follows, 

\begin{equation} P_n=C_n\hspace{2cm}n\leq 3 \end{equation}

\begin{equation} P_4=C_4+\frac{1}{7}\,\bar C_0\,(C_1^2-C_2\,C_0)
\end{equation}

\begin{equation} P_5=C_5+\frac{1}{3}\,\bar
C_0\,(C_2\,C_1-C_3\,C_0)+\frac{1}{21}\bar C_1\,(C_1^2-C_2\, C_0).
\end{equation}

As a function of these multipole moments the first coefficients in the
expansion  of the energy per unit of mass, $E$,
read,

\begin{eqnarray} E_1&=&0 \end{eqnarray}

\begin{eqnarray} E_2&=&-6-10\,E_0-{\frac {E_0^{2}}{2}} \end{eqnarray}


\begin{eqnarray} E_3&=&-\left({8}+{12\,E_{0}}\right)\frac{i\,P_{1}}{{ P_0}^{2}} \end{eqnarray}


\begin{eqnarray} E_4&=&-\frac {47}{4}-20\,E_0 -13\,E_0^{2}+
    \frac {E_0^{3}}{2} +(2+3\,E_0)\frac { P_2}{ {P_0}^{3}}
\end{eqnarray}


\begin{eqnarray} E_5&=&-\left(56+ 104\,E_0+56\,E_0^{2}
    \right)\frac{i\,P_{1}}{{ P_0}^{2}},
\end{eqnarray}
where $E_0$, the Keplerian energy, is a free parameter which has $-1/2$ as
a  lower bound, corresponding to a circular orbit. 

The frequency $\omega$ is different from one and therefore the orbit is not
closed. Between two  consecutives perihelion approaches the test particle
traces an angle  $2\,\pi/\omega$. Hence the perihelion has shifted an angle,
$\Delta\phi$  given up to the eleventh power of $\epsilon$ by the following
expression, 

\begin{eqnarray} {\Delta\phi}&=&2\,\pi\,(\omega^{-1}-1)=\pi\left\{
\Delta_{0}+\Delta_{1}+\Delta_{2}+\Delta_{4}+
\Delta_{8}+\right.
\nonumber\\&+&\left.\Delta_{16}+\Delta_{32}+\Delta_{2\times 4}+\Delta_{2\times 8}+\Delta_{2\times 16}+
\Delta_{4\times 8} \right\}, \end{eqnarray} 
where the shift has been split into different terms according to their 
origin: The first one, $\Delta_{0}$, comprises the Newtonian
contribution  to the precession, that is, the terms which remain after
taking the classical  limit $c\rightarrow\infty$. Of course only the
gravitational moments are  present since rotation has no influence
whatsoever in Newtonian dynamics.  Since the speed of light, $c$, and the
gravitational coupling constant, $G$,  have been taken to be one, the terms
look rather alike in magnitude. However,  if the respective factors are
written (a factor $G/c$ for each $\epsilon$,  a factor $G/c^2$ for each
$P_{2n}$, a factor $G/c^3$ for each $P_{2n+1}$,  a factor $c^{-1}$ for each
$l$ and a factor $G^{-2}$ for each $E_0$), the actual magnitude of every
term is recovered.  For instance, the first Newtonian term has a factor
$G^2$ and the second a  $G^4$. For oblate gravitational sources the
quadrupole moment, $P_2$ is  negative, whence it contributes to a positive
shift of the perihelion in  the first order. In the next order the shift
contribution of the quadrupole  is, however, always positive. (N.b. $E_0$,
though negative for  bounded orbits, has a lower limit about $-1/2$ which
does not allow it to overcome the energy-independent term.)  On the other
hand the sedecimpole $P_4$ term bears the opposite sign to  the one of the
quadrupole,


\begin{eqnarray}
    \Delta_{0}=-{\frac {3\,\,{ P_2}}{{
P_0}^{3}}}\,\epsilon^4+\left\{\left(\frac {105}{8}+\frac {45\,{ E_0}}{4}\right)
\frac{{ P_4}}{{ P_0}^{5}}+\left(\frac
{105}{8}+ \frac{15\,{ E_0}}{4}\right)\frac{{ P_2}^{2}}{{ 
P_0}^{6}}\right\}\epsilon^8.
\end{eqnarray}

The second term, $\Delta_{1}$, comprises the contribution to the
perihelion  precession due to a spherically symmetric mass distribution,
i.e. the  Schwarzschild effect. It can be calculated exactly in terms of
elliptic  functions and is of the order of $G^2/c^2$. The contribution of
every order  is always positive,

\begin{eqnarray}
\Delta_{1}&=&6\,\,\epsilon^{2}+\left ({\frac {
105\,}{2}}+15\,\,E_0\right )\epsilon^{4}+\left ({ \frac {975\,}{2}}+165\,\,E_0\right
)\epsilon^{6}+\nonumber\\&+&\left ({\frac {159105\, }{32}}+{\frac {16725\,\,E_0}{8}}+{\frac
{705\,\,E_0^{2}}{8}}\right )\epsilon^{8}+\nonumber\\&+&\left
({\frac {1701507\,}{32}}+ {\frac
{216375\,\,E_0}{8}}+{\frac {20115\,\,E_0^{2}}{8}}\right
)\epsilon^{10}.\label{permon} 
\end{eqnarray}

 In the term $\Delta_{2}$ we have included the influence of a dipole of 
 rotation on the perihelion shift. It is of the order of $G^2/c^2$. Since 
 its lower terms in $\epsilon$ are odd, it is sensitive to whether the probe
 rotates in the same direction as the source does or not. It is  positive if
 the angular momentum of the source and the orbital angular momentum of the
 probe are antiparallel and negative otherwise.  This has a qualitative
 explanation in the fact that the perihelion shift  due to a mass monopole
 decreases with the angular momentum of the test particle; if the source
 happens to be rotating and its angular momentum is  $J$, then $l$ is
 replaced in the first order by $l+2\,J/r$ as it can be seen  in equation
 (\ref{binet1}) after substitution of $P_1$ by $i\,J$, whence the 
 `effective' $l$ increases if both momenta are parallel and it would be 
 expected that the perihelion advance diminishes. In contrast, the quadratic 
 terms in $P_1$ are independent of the direction of rotation and are always
 positive whence they induce a perihelion advance. The cubic terms in  $P_1$
 behave as the linear ones,

\begin{eqnarray} 
\Delta_{2}&=&{\frac {8\,i\,{ P_1} \epsilon^{3}}{{
P_0}^{2}}}+
\left (168+48\,E_0\right )\frac{i\,P_{1}}{{ P_0}^{2}}\epsilon^{5}-
\left (120+24\,E_0\right )\frac{{ P_1}^{2}}{{ P_0}^{4}}\epsilon^{6}+
\nonumber\\&+&
\left (2562+ 1020\,E_0+36\,E_0^{2}\right
)\frac{i\,P_{1}}{{ P_0}^{2}}\epsilon^{7}
- \nonumber\\ &-&\left (\frac {65607\,}{16}
+\frac {44607\,E_0}{28
}+\frac {195\,E_0^{2}}{4}\right
)\frac{{ P_1}^{2}}{{ P_0}^{4}}\epsilon^{8}+\nonumber\\ &+&
\left \{\left(36046+17640\,E_0+1356\,E_0^{2}-16\,E_0^{3}\right)
\frac {i\, P_1}{{ P_0}^{2}}-\left(2048+672\,E_0\right)
\frac {i\,{ P_1}^ {3}}{{ P_0}^{6}}
\right
\}\epsilon^{9}-\nonumber\\&-&
\left (\frac {10256685}{112}+\frac {1320387\,E_0
}{28}+\frac {118305\,E_0^{2}}{28}\right )\frac{{ P_1}^{2}}{{ P_0}^{4}}\epsilon^{10}+\nonumber\\&+&
\left \{\left( \frac {3927489}{8}+\frac
 {569361\,E_0}{2}+\frac {70659
\,E_0^{2}}{2}+174\,E_{0}^3+15\,E_0^{4}
 \right)\frac{i\,{ P_1}}{{P_0}^{2}}-\right.
 \nonumber\\&-& \left.\left(\frac {2735961}{28}+\frac {341339\,E_0}
 {7}+\frac {27429\,
E_0^{2}}{7}\right )\frac{i\,{P_{1}}^3}{{ P_0}^{6}}\right\}\epsilon^{11}. 
\end{eqnarray}

Under the name $\Delta_{4}$ the relativistic terms depending only on the 
quadrupole moment, $P_2$, and the mass are comprised. The first correction 
has a factor $G^4/c^2$ in front of it. As it was to be expected, it does
not  depend on the direction of rotation and, as its Newtonian counterpart,
it is  positive for oblate gravitational sources. The quadratic terms as a
whole are  always positive,

\begin{eqnarray} \Delta_{4}&=&-\left (90 
+42\,E_0\right )\frac{{ P_2}}{{P_0}^{3}}\epsilon^{6}-\left
({\frac
{25383\,}{16}}+{\frac {28305\,\,E_0}{ 28}}+
{\frac {375\,E_0^{2}}{4}} \right )\frac{{ P_2}}{{ P_0}^{3}}\epsilon^{8}+
\nonumber\\&+&\left\{-\left ({\frac
{ 2686203}{112}}+{\frac {503379\,E_0}{28
}}+{\frac {80187\,E_0^{2}}{28}}\right)\frac{{ P_2}}{{ 
P_0}^{3}}+\right.\nonumber\\ &+&\left.\left({\frac
{12471}{16}}+519\,E_0
+{\frac {165\,E_ 0^{2}}{4}}\right
)\frac{{ P_2}^{2}}{{ P_0}^{6}}\right\}\epsilon^{10} 
\end{eqnarray}

The symbol $\Delta_{8}$ stands for the corrections to the perihelion
shift due to a rotation octupole moment $P_3$. The first correction is of
the order of $G^4/c^2$. They bear the same relation to the dipole terms as
the quadrupole to the monopole terms: The overall sign changes,


\begin{eqnarray} \Delta_{8}&=& -\left\{ \left ({30}+{24\,E_0}\right )\epsilon^{7}
    +\left
({801}+{825\,E_0}+{138\,E_0^{2}
}\right )\epsilon^{9}+\nonumber\right.\\ &+&\left. \left ({\frac
{28671}{2}}+{16475\,E_0}+{4278\,E_0^{2}}+{102\,E_0^{3}
}\right )\epsilon^{11} 
\right\} \frac{i\,P_{3}}{{P_{0}}^4}
\end{eqnarray}

The influence of the sedecimpole gravitational moment, $P_4$, is included 
in $\Delta_{16}$ and does not show up until the tenth order of the small 
parameter. In non-geometrized units it is proportional to $G^6/c^2$,
\begin{eqnarray} 
\Delta_{16}=\left ({\frac {7425}{16}}+{
570\,E_0}+{\frac {495\,E_0^{2}}{4}}\right )\frac{{ P_4}}{{
P_0}^{5}}\epsilon^{10}.
\end{eqnarray}


The last multipole moment to be considered is the rotational
trigintaduopole  moment, $P_5$, and it is comprised in $\Delta_{32}$. It
is of the eleventh order in $\epsilon$,

\begin{eqnarray} 
\Delta_{32}= \left ({\frac {945}{8}}+{
210\,E_0}+{\frac {135\, E_0^{2}}{2}}\right )\frac{i{ P_5}}{{
P_0}^{6}}\epsilon^{11}.
\end{eqnarray}

Now we review the couplings among the different multipole moments other
than  mass. Up to the order considered there is no coupling between the 
gravitational moments higher than the mass (except for self-couplings), but 
there are rotation-rotation couplings and gravitation-rotation couplings.
The  first one to appear is the dipole-quadrupole coupling, $\Delta_{2\times 4}$. 
It has a factor of $G^4/c^2$ in the lowest order. If both angular momenta
are  antiparallel and the source is oblate ($P_2<0$), then the contribution
of the bilinear terms is positive. The quadratic terms in the quadrupole
$P_2$ are  again positive if $J$ and $l$ are antiparallel. Finally the
quadratic terms  in the dipole are positive if the body is oblate,

\begin{eqnarray} 
\Delta_{2\times 4}&=& -\left (90+24\,E_0\right
)\frac{\,i{ P_1}\,{ P_2}}{{ P_0}^{5}}\epsilon^{7}-
\left (3939+2127\,E_0+126\,E_0^{2}\right
)\frac{\,i{ P_1}\,{ P_2}}{{ P_0}^{5}}\epsilon^{9}+
\nonumber\\&+&
\left (2280+900\,E_0 \right)\frac{{ P_1}^{2}{ P_2}}{{ P_0}^{7}}\epsilon^{10}
+
\left\{\left ({\frac {1419}{2}}+303\,E_{0}\right)\frac{i\,{ P_1}{ P_2 }^{2}}
{{ P_0}^{8}}-
\right.\nonumber\\&-&\left.
\left({\frac
{2825301 }{28}}+{\frac {507992\,E_0}{7
}}+{\frac {75765\,E_0^{2}} {7}}+90\,E_0^{3}\right )
\frac {i\,{ P_1} { P_2}}{{
P_0}^{5}}\right\}\epsilon^{11}.
\end{eqnarray}

The only rotation-rotation coupling up to this order is between the dipole 
and octupole moments. It comes under the name of $\Delta_{2\times 8}$ and it is
at  least of the order $G^6/c^4$. Since it is linear on both rotational
moments  this term is not sensitive to the direction of rotation. It is of
the tenth order in $\epsilon$,


\begin{eqnarray}
\Delta_{2\times 8}=\left ({1068} +
{972\,\,E_0}+{120\,E_0^{2}}\right )
\frac{{ P_1}\,{ P_3}}{{P_0}^{6}}\epsilon^{10}.
\end{eqnarray}

The higher rotation-gravitation couplings involve the rotational dipole and
the gravitational sedecimpole, $\Delta_{2\times 16}$, and the rotational octupole
and the gravitational quadrupole, $\Delta_{4\times 8}$. Both appear first in the
eleventh order of perturbation and are at least of the order of
$\epsilon^{11}$. The term $\Delta_{2\times 16}$ is positive if $P_4$ is positive
and the angular momenta are antiparallel,

 \begin{eqnarray} 
 \Delta_{2\times 16}= \left ({\frac {4005}{8}}+{ {495\,E_0}}+{\frac {
 135\,E_0^{2}}{2\,}}\right )\frac{i{ P_1}\,{ P_4}}{{ P_0 }^{7}}\epsilon^{11}
 \end{eqnarray}

  \begin{eqnarray} 
 \Delta_{4\times 8}= \left ({\frac {1383}{4
 }}+{348\,E_0}+{
 45\,E_0 ^{2}}\right )
 \frac{i{ P_2}\,{ P_3}}{{ P_0}^{7}}\epsilon^{11}.
 \end{eqnarray}


It would be of great interest to know the range of applicability of this 
perturbative expansion. In an appendix at the end of this paper a simpler 
case  is studied: It is shown there for which values of the parameters the 
expansions are acceptable for the Schwarzschild metric. 

\section[Precession of the line of nodes]{Precession of the line of nodes of a nonequatorial
orbit\label{node}} 

Let us consider now a bounded orbit slightly departing from the equatorial 
plane. If the mass distribution of the source were spherically symmetric, 
the equatorial plane would not be at all privileged and the orbit would 
always intersect it at the same nodes. However this is no longer the 
situation when the source is not exactly spherical, since then the nodes 
precess due to the perturbation generated by higher order multipoles. In 
classical nonrelativistic mechanics the first contribution to the
precession  of the nodes arises from the quadrupole moment, as it is shown
in an  appendix at the end of this paper, whereas in general relativity it
is the rotational dipole moment the first one to contribute.

In the approximation used in this section we consider a geodesic on the 
equatorial plane and calculate the evolution of small deviations from it. 
Since the geodesic equation reads,

\begin{equation} \ddot x^\mu+\Gamma^\mu_{\rho\sigma}\,\dot x^\rho\,\dot
x^\sigma=0, \end{equation} 
it is straightforward that nearby geodesics which deviate from the original 
geodesic by a vector $\delta^\mu$ fulfill,

\begin{equation}
\ddot\delta^\mu+2\,\Gamma^\mu_{\rho\sigma}\,\dot\delta^\rho\,\dot
x^\sigma+\Gamma^\mu_{\rho\sigma,\nu}\,\dot x^\rho\,\dot
x^\sigma\,\delta^\nu=0. \end{equation}

Since we are interested in small deviations from the equatorial plane,  we
shall focus our attention on the $\theta$ coordinate.  Taking into account
that the reference geodesic lies on the symmetry plane  $\theta=\pi/2$ and
that the first derivatives of the metric with respect  to $\theta$ vanish
on it, the geodesic deviation equation for  $\delta^\theta$ reduces to,

\begin{equation}
\ddot\delta^\theta-\frac{1}{2}\,g^{\theta\theta}\,g_{\rho\sigma,\theta\theta}
\,\dot x^\rho\,\dot x^\sigma\,\delta^\theta=0. \end{equation}

Instead of considering an arbitrary bounded reference geodesic on the 
equatorial plane, we shall restrict ourselves to geodesic circles. Of 
course many interesting features will be lost, but we deem this paper 
would become even longer if we take them into account.

In order to compute the evolution of the nodes with respect to the
azimuthal  angle on the geodesic, the previous equation needs to be divided
by  $\dot\phi^2$, which is constant on the circle. From now on we shall
write  $\delta$ instead of $\delta^\theta$ to avoid cumbersome 
notations, 

\begin{equation}
\delta_{\phi\phi}+\Omega^2\,\delta=0\hspace{1.5cm}\Omega^2=-\frac{1}{2\,
\dot\phi^2 }\, g^{\theta\theta}\,g_{\rho\sigma,\theta\theta}\,\dot
x^\rho\,\dot x^\sigma, \end{equation} 
where every function is to be calculated on $\theta=\pi/2$ and $r=R$,  the
radius of the geodesic circle. 

The previous expression states that the nodes of nearby geodesics are 
separated by regular intervals of the coordinate $\phi$. If $\Omega$ is 
different from one, these nodes will travel around the geodesic circle 
instead of remaining at constant values of the azimuthal angle. Since it 
would be of interest to write down the result in a coordinate independent 
expression, it will be required to make use of the geodesic equations to 
remove the dependence on the radius of the circle, $R$, and also to cast 
the energy, $E$, as a function of the angular momentum of the circular 
orbit, $l$, so that the final result be a function of the multipole moments 
and $l$ only.

As it is well known, the timelike geodesic equations can be obtained from 
the Lagrange equations applied to a functional  $L=g_{\mu\nu}\,\dot
x^\mu\,\dot x^\nu$ and the constraint $L=-1$.  Two of these equations,
namely (\ref{phi}) and (\ref{t}), have already  been used to remove the
dependence on the derivatives of the ignorable  coordinates. The equation
corresponding to variations of $\theta$ is  automatically satisfied. We are
left just with two equations, corresponding  to the constraint and the
variations of $r$, viz. 

\begin{equation} 1=\frac{E^2}{f}-f\,\frac{(l-E\,A)^2}{r^2}\label{tim}
\end{equation}

\begin{equation}
\frac{E^2}{f^2}\,\partial_rf-2\,E\,f\frac{l-E\,A}{R^2}\partial_rA-f^2
\frac{(l-E\,A)^2}{R^4}\partial_r\left(\frac{r^2}{f}\right)=0, \end{equation} 
where we have used $\dot r=0$ and $\theta=\pi/2$. 

From the last equation we get $E/l$ as a function of $R$, 

\begin{equation} \frac{E}{l}=\frac{-b\pm\sqrt{b^2-4\,a\,c}}{2\,a},
\end{equation}

\begin{eqnarray}
a&=&A^2\,f^2\,R^{-4}\,\partial_r(r^2\,f^{-1})-f^{-2}\,\partial_rf-2\,A\,f\,
R^{-2}\partial_rA=\nonumber\\&=&-2\,P_0\,R^{-2}+O(R^{-3}) \end{eqnarray}

\begin{eqnarray}
b&=&2\,R^{-2}\,f\partial_rA-2\,A\,f^2\,R^{-4}\partial_r(r^2\,f^{-1})=\nonumber\\
&=&-12\,i\,P_1\, R^{-4}+O(R^{-5}) \end{eqnarray}

\begin{eqnarray} c=f^2\,R^{-4}\partial_r(r^2\,f^{-1})=2\,R^{-3}+O(R^{-4}).
\end{eqnarray}

Since the energy must be positive, the solution with the minus (plus)
signus  in front of the square root equation corresponds to a positive
(negative)  $l$. On the other hand equation (\ref{tim}) furnishes $l$ in
terms of $E/l$ which has obviously the same coefficients which were obtained
for $E$ in  the previous section after substitution of $E_0=-1/2$. 

The frequency of the precession of the nodes can be written now
independently  of the choice of coordinates. As was to be expected, the
first term is  equal to one, corresponding to the frequency of the nodes
when the mass  distribution of the gravitational source is perfectly
spherical. 

The contributions to the precession of the line of nodes of a timelike 
geodesic which is slightly tilted with respect to a geodesic equatorial
circle  have been classified in the same way as it was done for the
perihelion shift:  A classical term plus the relativistic terms, divided
according to their  multipole content. In order to write down the
expressions in non-geometrized  units,  factors including $G$ and $c$ have
to be included as it was done in the  previous section. The description
which was made there of the necessary  factors for each term in
$\Delta\phi$ for the perihelion shift also  applies. The information that
we have of the metric allows us to calculate the coefficients up to
$l^{-13}$.

There is, of course, no contribution from a pure mass monopole. Were the 
frequency equal to one, then the nodes would remain at constant $\phi$. 
Therefore, $\Delta\phi=2\,\pi\,(\Omega^{-1}-1)$  describes the angle
through which the line of nodes has precessed in one  revolution  of the
reference circle,

\begin{eqnarray} \Delta\phi&=&\pi\,\left\{\Delta_{0}
+\Delta_{2}+\Delta_{4}
+\Delta_{8}+\Delta_{16}+\Delta_{32}+\Delta_{2\times 4} +\right.
\nonumber\\&+&\left.\Delta_{2\times 8}+\Delta_{2\times 16}+\Delta_{4\times 8} \right\}
\end{eqnarray}

For oblate objects ($P_2<0$), the influence of the mass quadrupole amounts 
to a delay in the precession of the line of nodes with respect to the
$\phi$  coordinate on the circle of reference. Therefore the line of nodes
does not  precess in the same direction as the perihelion. This fact should
not be  confused with the precession of the angular momentum vector in
time, which  is of course positive. The contribution  from the sedecimpole
term is negative for positive $P_4$ and the one from the  sexagintaduopole
is positive for positive $P_6$. There is also a classical  coupling between
$P_4$ and $P_2$. The nonlinear terms in the quadrupole  moment bear a
positive sign,

\begin{eqnarray} 
\Delta_{0}&=&\frac{3\,{ P_0}\,{ P_2}}{l^{4}}+\left
(-{\frac {15\,{ P_0}^{3}{
P_4}}{2}}+{\frac {9\,{ P_0}^{2}{ P_2}^{2}}{4}}\right )l^{-8}+\nonumber \\&+&
\left ({\frac {105\,{ P_0}^{5}{ P_6}}{8}}+{\frac {45\,{ P_0}^{4}{ P_2 }\,{
P_4}}{8}}+
{\frac {81\,{ P_0}^{3}{ P_2}^{3}}{8}}\right )l^{-12}.
\end{eqnarray}

Most of the dipole-dependent terms are sensitive to the direction of 
motion of the test particle  relative to the rotation of the source. If the
angular momenta are parallel  ($l$ has the same sign as $J=-i\,P_1$), then
the linear and cubic terms in  $P_1$ induce an advance of the line of
nodes. On the other hand the  quadratic and quartic terms in the dipole
moment are not sensitive to that  relative sign and their influence always
amounts to a delay,

\begin{eqnarray} 
\Delta_{2}&=&-\frac{4\, i\,{ P_0}{
P_1}}{l^{3}}-\frac{18\,i\,{ P_0}^{3}{ P_1} }{l^{5}}+\frac{18\,{ P_0}^{2}{
P_1}^{2}}{l^{6}}-\frac {243\,i\,{ P_0}^{5}{ P_1}}{2\,l^7}+\frac
{4131\,{ P_0}^{4}{ P_1}^{2}}{14\,l^8}+\nonumber\\&+&\left (196\,i{ P_0}^{3} { P_1}^{3}-{
\frac {3861\,i{ P_0}^{7}{ P_1}}{4}}\right)l^{-9} +\frac
{27294 \,{ P_0}^{6}{ P_1}^{2}}{7}\,l^{-10}+ \nonumber\\&+&\left ({\frac { 37983\,i{
P_0}^{5}{ P_1}^{3}}{7}}-{\frac { 268515\,i{ P_0}^{9}{ P_1}}{32}} \right
)l^{-11}+\nonumber\\&+&\left (-2565\,{ P_0}^{4}{ P_1}^{4}+{\frac
{1060043\,{ P_0}^{8}{ P_1}^{2}}{22}}\right )l^{-12} +\nonumber\\&+& \left
({\frac {734200\,i{ P_0}^{7}{ P_1}^{3}}{7}}-{\frac { 4944807\,i{ P_0}^{11}{
P_1}}{64}} \right )l^{-13}.
\end{eqnarray}

The relativistic contribution of the linear terms in the quadrupole moment
$P_2$ is again negative for oblate sources. The quadratic terms furnish a
negative contribution to the precession shift regardless of whether the
source is oblate or prolate, whereas the classical quadratic correction is
positive, as it was shown previously,

\begin{eqnarray} 
\Delta_{4}&=&\frac{24\,{ P_0}^{3}{ P_2}}{l^{6}}+\frac
{2799\,{ P_0}^{5}{ P_2}}{14\,l^{8}}+\left (-24\,{ P_0}^{4}{
P_2}^{2}+{\frac {12396\,{ P_0}^{7}{
P_2}}{7}}\right)l^{-10}+\nonumber\\&+& \left (-{\frac
{327223\,{P_0}^{6}{P_2}^{2}}{308}}+ {\frac
{361993\,{ P_0}^{9}{ P_2}}{22}}\right )l^{-12}.
\end{eqnarray}

As it happened with the perihelion precession, the rotational octupole term 
has the  opposite sign to the one of the linear dipole term and of course
it is  dependent  on the direction in which the probe orbits,

\begin{eqnarray} 
\Delta_{8}&=&\frac{12\,i{ P_0}^{3}{ P_3}}{
l^{7}}+\frac{156\,i{ P_0}^{5}{ P_3}}{l^{9}}+{\frac {3387\,i{ P_0}^{7}{
P_3}}{2 \,l^{11}}}+\frac{17707\,i{ P_0}^{9} { P_3}}{l^{13}}.
\end{eqnarray}

The relativistic sedecimpole term bears the same sign as its classical 
counterpart, although it is much smaller in magnitude. There is no 
relativistic term in $P_6$,  just the classical term which has already been
described,

\begin{eqnarray} 
\Delta_{16}=-\frac{120\,{ P_0}^{5}{ P_4}}{l^{10}}-{\frac
{2905\,{ P_0}^{7}{ P_4}}{2\,l^{12}}},
\end{eqnarray}
and again we have another permutation of sign; the term in $P_{5}$ 
bears the same sign as the linear dipole term,

\begin{eqnarray} 
\Delta_{32}= -{\frac {45\,i{ P_0}^{5}{ P_5}}{2\,l^{11}}}
-{\frac {1905\,i{ P_0}^{7} { P_5}}{4\,l^{13}}}.
\end{eqnarray}

Now we discuss the coupling terms between multipole moments other than the 
mass. The bilinear coupling between the dipole and the quadrupole moment 
is positive for oblate gravitational sources which rotate in the same 
direction as the test particle. The quadratic term in $P_1$ is negative 
for oblate objects no matter in which direction they rotate. The quadratic 
term in $P_2$ is positive when the angular momenta $J$ and $l$ are
parallel,

\begin{eqnarray} 
\Delta_{2\times 4}&=&\frac{18\,i{ P_0}^{2}{ P_1}\,{ P_2}}{l^{7}}
+\frac{441\,i{ P_0}^{4} { P_1}\,{ P_2}}{l^{9}}-\frac{273\,{ P_0}^{3}{
P_1}^{2}{  P_2}}{l^{10}}+\nonumber\\&+& \left (-{\frac { 81\,i{ P_0}^{3}{
P_1}\,{ P_2}^{2}}{2}}+{ \frac {198369\,i{ P_0}^{6}{ P_1}\,{ P_2}}{28}
}\right )l^{-11}- {\frac {3007003\,{ P_0}^{5 }{ P_1}^{2}{
P_2}}{308\,l^{12}}}+ \nonumber\\&+&\left (-{\frac {10899i
{ P_0}^{5}{ P_1}{ P_2}^{2}}{4}}-4482i{ P_0}^{4 }{ P_1}^{3}{ P_2}+
{\frac {5509583i{ P_0}^{8}{  P_1}{
P_2}}{56}} \right
)l^{-13}.
\end{eqnarray}

The rotational bilinear coupling between the dipole and the octupole moment 
is independent of the direction of rotation and it is positive when 
$J=-i\,P_1$ and $J_3=-i\,P_3$ have the same sign. There is however a 
higher coupling which is quadratic in the dipole moment and contributes  in
the same way as $\Delta_{8}$ does,

\begin{eqnarray} 
\Delta_{2\times 8}=-\frac{186\,{ P_0}^{4}{ P_1}\,{
P_3}}{l^{10}}-{\frac {100201\,{ P_0}^{6}{ P_1}\,{
P_3}}{22\,l^{12}}}-\frac{3168\,i{ P_0}^{5} { P_1}^{2}{ P_3}}{l^{13}}.
\end{eqnarray}

The last terms to be considered up to this order are the  couplings between
$P_1$ and $P_4$ and between $P_2$ and $P_3$, which are both sensitive to
the relative directions of rotation of the  source and the probe particle,

\begin{eqnarray} 
\Delta_{2\times 16}&=&- {\frac {255\,i{ P_0}^{4}{ P_1}\,{
P_4}}{2\,l^{11}}}- {\frac { 14193\,i{ P_0}^{6}{ P_1}\,{ P_4}}{4\,l^{13}}}
\end{eqnarray}

\begin{eqnarray} 
\Delta_{4\times 8}&=&- \frac{21\,i{ P_0}^{4}{ P_2}\,{
P_3}}{l^{11}}- {\frac {1971 \,i{ P_0}^{6}{ P_2}\,{ P_3}}{2\,l^{13}}}.
\end{eqnarray}

\section{Conclusions\label{discussion}}

In this paper we have displayed the approximate general asymptotically flat 
stationary axisymmetric metric for the vacuum spacetime surrounding a
compact  source possessing a symmetry plane orthogonal to the symmetry
axis.  The calculations have been carried out for the Ernst potential up to
the  term in $r^{-7}$ in the pseudospherical radial coordinate. This has
been  useful to calculate relativistic corrections to the classical orbits
around  a compact mass distribution. In particular the perihelion
precession on the  symmetry plane and the precession of the nodes of a
slightly tilted circular  orbit  have been calculated. 

Concerning the perihelion shift, it has been shown that the contributions
of  each gravitational and rotational multipole moment follow a curious
pattern  of alternation of signs: The term in the mass monopole is always
positive,  whereas the linear quadrupole term is negative for positive
$P_2$  (the quadratic term in $P_2$ is always positive) and the sedecimpole
term  is again positive for positive $P_4$. The sexagintaquattuorpole $P_6$
is outside the limits of our perturbative expansion. If we think for 
instance of an oblate source which is very close to a classical
axisymmetric  homogeneous ellipsoid, then we would also have alternating
signs in the  gravitational multipole expansion 
($P_{2\,n}=3\,P_0\,(c^2-a^2)^n/ [(2\,n+1)(2\,n+3)])$, $c$ and $a$ being the 
lengths of the ellipsoid's semiaxis parallel and orthogonal to  the
symmetry axis, respectively) and every term would have a positive 
contribution. The  linear and cubic rotational dipole term is always
positive for counterrotating  configurations of the source and the probe
(the quadratic  contribution is always positive regardless of the relative
rotation)  whereas the octupole term is positive if $J_3=-i\,P_3$ and $l$
bear the same sign. The trigintaduopole term has the same sign as  the
dipole term. It should be pointed out that the energy dependent  terms are
not strong enough to affect these signs. The coupling  between different
multipoles preserves the sign of the product of the  corresponding linear
terms, that is, if there is a $P_n-P_m$ coupling,  then its sign is
obtained from the product of the $-1$ factors multiplying the corresponding
linear terms in $P_n$ and $P_m$.  No coupling between mass multipole
moments of order higher than the monopole  appears to this order except for
self-couplings. The sign independent terms,  such as the mass monopole term
and the quadratic dipole and quadratic  quadrupole terms, always give rise
to a perihelion advance.

The influence of the different multipole moments on the precession  of the
line of nodes of a particle departing by a small amount from  the
equatorial plane is somewhat different from the behaviour which has  been
shown for the perihelion precession: The alternating pattern of the  signs
of the linear terms in the multipole moments other than the mass is 
preserved both for gravitational and rotational moments independently up 
to the considered perturbation order: If the $P_{2n}$ were all positive, 
then the contribution of each $P_{2n}$ would be opposite to the one of
$P_{2n+2}$ and a similar reasoning is valid for the rotational terms. 
However the relativistic coupling terms bear the opposite sign to the one 
which would be expected from the product of the $-1$ factors before the 
corresponding linear terms: For instance, if the $P_1$ and $P_2$ linear 
terms are positive then the corresponding bilinear coupling is negative. 
On the other hand, this rule is not valid for the Newtonian self-couplings.
Another difference arises from the fact that the gravitational 
sexagintaduopole moment $P_6$ does influence the precession of the line of 
nodes in this order of perturbation, whereas it does not affect the 
perihelion precession. Also, there were no couplings between the 
gravitational moments (self-couplings and mass-couplings excluded) in the 
perihelion precession, but there is one (the classical $P_2-P_4$-coupling) 
in the node precession. It is curious that, while the Newtonian 
sign-independent terms are always positive, the relativistic ones  (the
quadratic dipole and the quadratic quadrupole self-couplings) are 
negative. 

Of course most of these corrections are meaningless for astronomical 
purposes in our solar system, but are very likely to be relevant for 
highly relativistic astrophysical objects, suchs as pulsars and blackholes, 
where other post-Newtonian effects \cite{Schafer} have been shown to be
present. 

\appendix

\section{Schwarschild spacetime}

\noindent In this appendix we shall study the range of applicability of the 
perturbation expansion for bounded orbits in Schwarschild's spacetime. For 
this metric the geodesic equation (\ref{binet1}) can be solved exactly in 
terms of elliptic functions. Instead of writing it as a differential 
equation for $u$ as a function of $\phi$, we are writing it as a
differential  equation for $\phi$. In this section $u$ is no longer the
inverse of the  pseudospherical radius but of the usual Boyer-Lindquist
radius,

\begin{equation}
\phi_u=\left(2\,P_0\,u^3-u^2+\frac{2\,P_0}{l^2}\,u+\frac{E^2-1}{l^2}\right)^
{-1/2}=g(u)^{-1/2}.\label{binetsch} \end{equation}

We are considering that $E<1$ and therefore $g(u)$ has at least one zero. 
For bounded motion we need three zeros so that the orbit ranges between the 
apsidal points. If we call these zeros $a\geq b\geq c\geq 0$, then equation 
(\ref{binetsch}) can be integrated \cite{tablas} in terms of the elliptic 
integral of the first kind, $F(\gamma,q)$, in the region $b\geq u>c$, 

\begin{equation}
(\phi-\phi_0)\,\sqrt{\frac{P_0\,(a-c)}{2}}=F(\gamma,q)=\int_0^\gamma
d\alpha\, (1-q^2\,\sin^2\alpha)^{-1/2}\end{equation}

\begin{equation} \gamma=\arcsin\sqrt{\frac{u-c}{b-c}}\hspace{1.5cm}q=
\sqrt{\frac{b-c}{a-c}}. \end{equation}

An expression for $u$ in terms of $\phi$ is easily obtained taking into 
account that the elliptic sine is the sine of $F(\gamma,q)$,

\begin{equation} u=c+(b-c)\textrm{sn}^2\left\{\sqrt{\frac{P_0\,(a-c)}{2}}
\,(\phi-\phi_0)\right\}. \end{equation}

Since the real period of the elliptic sine is $4\,K(q)=4\,F(\pi/2,q)$, then
our $u$ function is $2\,K(q)$-periodic. Therefore the exact perihelion
precession of the orbit of a test particle around a spherical non-rotating
compact object will be,

\begin{equation}
\Delta\phi=\frac{2\,\sqrt{2}\,K(q)}{\sqrt{P_0\,(a-c)}}-2\,\pi,
\end{equation} 
the perturbation expansion of which in  $\epsilon$ coincides with the one
of $\Delta_{1}$ in equation (\ref{permon}),  as it was to be 
expected,

\begin{equation}
K(q)=\frac{\pi}{2}\left\{1+\sum_{n=1}^\infty\left(\frac{(2\,n-1)!!}{2^n\,n!}
\right)^2\,q^{2\,n}\right\}. \end{equation}

The limits of the range of applicability of the previous expansion are 
$q=1$ ($a=b$) and $q=0$ ($b=c$).  For both values, there are two zeros of
$g(u)$ which coalesce and the  bounded motion is no longer stable.
Therefore we are led to study the range  of parameters for which $g(u)$ has
double roots. This happens when $u$  takes either of the values $u_\pm$
which are solutions of $g'(u)=0$,

\begin{equation} u_\pm=\frac{1\pm\sqrt{1-12\epsilon^2}}{6\,P_0},
\end{equation} 
and therefore the allowed region is the one enclosed between the  curves
$g(u_\pm)=0$ in the $E^2-q^2$ parameter plane. This yields critical values 
for the energy per unit of mass, $E_c^2=8/9$, and the perturbation
parameter,  $\epsilon_c^2=1/12$. The perturbative approach is no longer
valid beyond  this point in the parameter plane since there are no stable
bounded orbits.  It is remarkable that for $\epsilon^2>1/16$ there is not
only a lower limit  for the energy of the bounded orbit but also an upper
limit. 

\section{Classical precession of the line of nodes}

\noindent In this appendix we shall briefly derive the classical expression 
for the precession of the line of nodes.

The Lagrangian for the motion of a particle in a gravitational field 
is,

\begin{equation} L=\frac{1}{2}\,\dot r^2+\frac{1}{2}\,r^2\dot
\theta^2+\frac{1}{2}\, r^2\,\sin^2\theta\,\dot\phi^2-V(r,\theta)
\end{equation}

\begin{equation} V(r,\theta)=-\sum_{n=0}^\infty
\frac{P_n\,p_n(\cos\theta)}{r^{n+1}}. \end{equation}

The equations of motion and conserved quantities which can be obtained from 
this Lagrangian are:

\begin{equation} E=\frac{1}{2}\,\dot r^2+\frac{1}{2}\,r^2\dot
\theta^2+\frac{1}{2}\, r^2\,\sin^2\theta\,\dot\phi^2+V(r,\theta)
\end{equation}

\begin{equation} l=r^2\,\sin^2\theta\dot\phi \end{equation}

\begin{equation} \ddot
r=r\,\dot\theta^2+r\,\sin^2\theta\,\dot\phi^2-\partial_rV(r,\theta)
\end{equation}

\begin{equation} r^2\,\ddot\theta+2\,r\,\dot
r\,\dot\theta=r^2\,\sin\theta\,\cos\theta\,\dot\phi^2-\partial_\theta
V(r,\theta)\label{theta}. \end{equation}

Truncating the Legendre expansion at the sexagintaquattuorpole multipole
moment, $P_6$, we get the following expressions for the energy per unit of
mass, $E$, and the radius, $R$, of a circular orbit on the plane
$\theta=\pi/2$ in the far field region,

\begin{equation}
E=-\frac{P_0^2}{2\,l^2}+\frac{P_0^3\,P_2}{2\,l^6}-\frac{3\,P_0^5\,P_4+9\,
P_0^4\,P_2^2}{8\,l^{10}}+O(l^{-14}) \end{equation}

\begin{eqnarray} R^{-1}=\frac{P_0}{l^2}-\frac{3\,P_0^2\,P_2}{2\,l^6}+
\frac{15\,P_0^{4}\,P_4+36\,P_0^3\,P_2^2}{8\,l^{10}}+O(l^{-14}).
\label{radius} \end{eqnarray}

In order to obtain the oscillations about the equatorial plane of a
slightly tilted bounded trajectory with respect to the circular orbit, we
introduce a small variation in the equation (\ref{theta}). The result will
be divided by $\dot\phi^2$ to yield the evolution of $\delta\theta$ as a
function of $\phi$,

\begin{equation} (\delta\theta)_{\phi\phi}=-\Omega^2\,\delta\theta
\hspace{1.5cm}\Omega=\sqrt{1+\frac{R^2}{l^2}\,V_{\theta\theta}(R,\pi/2)}.
\end{equation}

From this expression we get the frequency of the oscillations, $\Omega$,
which  can be written in terms of $l$ and the multipole moments by
inserting equation  (\ref{radius}) into it,

\begin{eqnarray}
\Delta\phi&=&2\,\pi\,\left(\frac{1}{\Omega}-1\right)=\frac{3\,\pi\,P_0\,P_2}
{l^4}+\frac{\pi\,(9\,P_0^2\,P_2^2-30\,P_0^3\,P_4)}{4\,l^8}+\nonumber\\&+&\frac{\pi\,
(105\,P_0^5\,P_6+45\,P_0^4\,P_2\,P_4+81\,P_0^3\,P_2^3)}{8\,l^{12}}+O(l^{-16}),
\end{eqnarray} 
which obviously coincides with the term $\Delta_{0}$ which was 
calculated using the full relativistic theory in section \ref{node}. 

\vspace{4mm} \appendix{\noindent\large{\bf Acknowledgements}}  \vspace{4mm}

\noindent The present work has been supported by Direcci\'on General de
Ense\~nanza Superior Project PB98-0772. L.F.J. wishes to thank
 F.J. Chinea, L.M. Gonz\'alez-Romero, F. Navarro-L\'erida and  M.J. Pareja for valuable discussions. L.F.J. wishes to thank  the
Department of Mathematical Sciences of the Loughborough University of 
Technology for their hospitality.


\begin{thebibliography}{99} 
\bibitem{Dietz} {\it Solutions of Einstein's
equations: Techniques and Results} (eds.: C. Hoenselaers and W. Dietz,
Springer Verlag), Berlin-New York (1984) 
\bibitem{Geroch}R. Geroch, {\it J. Math. Phys.} {\bf 11} 2580 (1970) 
\bibitem{Hansen}R. O. Hansen, {\it J. Math. Phys.} {\bf 15} 46 (1974) 
\bibitem{Fodor}G. Fodor, C. Hoenselaers, Z.
Perj\'es, {\it J. Math. Phys.} {\bf 30} 2252 (1989) 
\bibitem{Schafer} T. D'amour, G. Sch\"afer, {\it Nuovo Cimento} {\bf B 101} 127 (1988)
\bibitem{quadr} C. Hoenselaers, {\it Prog. Theor. Phys.} {\bf 56} 324
(1977) 
\bibitem{Quevedo} H. Quevedo, {\it Ph. D. Thesis} Universit\"at zu
K\"oln (1987) \bibitem{Ernst}F. J. Ernst, {\it Phys.Rev.} {\bf 167} 1175
(1968) 
\bibitem{Goldstein}H. Goldstein, {\it Classical
Mechanics}, Addison-Wesley (1980) 
\bibitem{Perlick}V. Perlick, {\it Class. Quantum
Grav.} {\bf 9} 1009 (1992) 
\bibitem{tablas} I. S. Gradsteyn, I. M. Ryzhik,
{\it Table of integrals, series and products}, Academic Press, New York
(1965)  
\end{thebibliography}
\end{document}